\documentclass[page-classic]{epl2}
\usepackage{color}
\usepackage{amssymb}
\usepackage{multirow}
\usepackage{bigstrut}

\title{\bf Synchronization in Scale Free networks: The role of finite size effects}  

\author{D. Torres\inst{1} \and M. A. Di Muro\inst{1} \and C. E. La Rocca\inst{1} \and L. A. Braunstein\inst{1,2}}
\shortauthor{D. Torres \etal}

\institute{ \inst{1} Instituto de Investigaciones F\'isicas de Mar del
  Plata (IFIMAR)-Physics Department, Universidad Nacional de Mar del
  Plata-CONICET - Funes 3350, (7600) Mar del Plata,
  Argentina.\\ \inst{2}Center for Polymer Studies, Boston University -
  Boston, Massachusetts 02215, USA }


\abstract{Synchronization problems in complex networks are very often
  studied by researchers due to its many applications to various
  fields such as neurobiology, e-commerce and completion of tasks. In
  particular, Scale Free networks with degree distribution $P(k)\sim
  k^{-\lambda}$, are widely used in research since they are ubiquitous
  in nature and other real systems. In this paper we focus on the
  surface relaxation growth model in Scale Free networks with $2.5<
  \lambda <3$, and study the scaling behavior of the fluctuations, in
  the steady state, with the system size $N$. We find a novel behavior
  of the fluctuations characterized by a crossover between two regimes
  at a value of $N=N^*$ that depends on $\lambda$: a logarithmic
  regime, found in previous research, and a constant regime. We
  propose a function that describes this crossover, which is in very
  good agreement with the simulations.  We also find that, for a
  system size above $N^{*}$, the fluctuations decrease with $\lambda$,
  which means that the synchronization of the system improves as
  $\lambda$ increases. We explain this crossover analyzing the role of
  the network's heterogeneity produced by the system size $N$ and the
  exponent of the degree distribution.}

\begin{document}

\maketitle

Since a great variety of systems can be represented by complex
networks, over the last decades many researchers have studied both the
topology and processes that evolve on top of these networks. Systems
such as neural networks, the Internet and airlines networks
\cite{Barabasi,Dorogovtsev,Verma} can be described by a set of nodes
connected by links that represent a relationship between them, such as
an electric impulse, friendship or air traffic. Many of these real
networks were found to be characterized by a Scale Free (SF) topology,
given by a degree distribution
\begin{equation}
P(k)\sim k^{-\lambda} ,
\end{equation} 
where $k$ is the degree of the nodes and $m \leq k \leq k_{max}$,
where $m$ and $k_{max}$ are the minimum and maximum degree
respectively, and $ \lambda $ represents the broadness of the
distribution. On most real systems, such as the World Wide Web or
metabolic networks, it was found that $ 2<\lambda <3 $
\cite{Barabasi,Dorogovtsev}.

More recently, research has focused on dynamical processes taking
place on the underlying network
\cite{watts,newmansiam,Li,Lucas,Lidia,Motter,sincronizacion}. Particularly,
many mathematical and numerical models have been elaborated to study
the problem of synchronization
\cite{Arenas,Kornis,Ana,Cristian,Cristian2,Cristian3}, a phenomenon
present in the behavior of many collective systems. In these processes
the state of the system evolves to a synchronized state, where the
coupled units adjust their dynamics with one another. Examples of
synchronization can be seen in brain processes \cite{Xiao} or data
distribution \cite{Valverde,Rabani,Kornis2}; in a network made up of
processors that distribute the task load, the system is best
synchronized when the process minimizes the waiting time in each
processor.  For these kinds of systems, a scalar field $h$ is usually
defined on the network and it is of interest to measure the
fluctuations of $h$. This problem can be studied mapping it into a
non-equilibrium surface growth problem, where the scalar field $h_i$,
with $i=1,...N$ and $N$ is the system size, represents the ``height''
of the node $i$, and the fluctuations, also called roughness of the
system, are given by

\begin{equation}
W(t)=\sqrt{\frac{1}{N}\sum_{i=1}^N[h_i(t)-\langle h(t)\rangle]^{2}} \, ,
\end{equation}
and
\begin{equation}
\langle h(t)\rangle=\frac{1}{N}\sum_{i=1}^Nh_i(t) \, ,
\end{equation}
is the mean value of $h$ at time $t$. The roughness has two regimes,
one where $W(t)$ increases until it saturates at a constant value
$W_s$ in the second regime, which depends on the topology of the
system. In Euclidean lattices one of the most studied equations on
surface growth is the Edward Wilkinson (EW) equation \cite{EW}, which
belongs to the same universality class that the stochastic growth
model of surface relaxation to the minimum (SRM) \cite{family}. In
this model, a node of the network is randomly selected and a
``particle'' is placed on the node with the lowest height among the
selected node and its neighbors. According to this rule, the nodes
with higher heights distribute their excess of particles to their
neighbors with lower heights.  Pastore y Piontti $et$ $al.$ \cite{Ana}
studied the SRM model on SF networks through numerical simulations and
found that, on the steady state, the behavior of the fluctuations with
the system size $N$ is given by
\begin{equation}\label{ws_ana}
W_{s}\sim \left \{\begin{array}{ll}
  const. & \textrm{, for $ \lambda\geq 3 $}\, ;\\
  ln\,N & \textrm{, for $ \lambda< 3 $}\, .
\end{array}\right.
\end{equation}  
Unlike Euclidean lattices, in complex networks one cannot extend the
discrete nature of the network to a continuum, and thus the dynamics
of the network are not well represented by a continuum equation such
as the EW equation. However, using a discrete Laplacian and a mean
field approximation, Korniss $et$ $al.$ \cite{Kornis} and Guclu $et$
$al.$ \cite{Kornis3} found that in the limit $N \rightarrow \infty $,
$W_s$ increases with $\lambda$. Solving the discrete EW
equation numerically for finite size systems, in \cite{Ana} the
authors found that $W_s$ decreases with $N$, which is not
representative of any growth model. With a different approach, La
Rocca $et$ $al.$ \cite{Cristian} developed a Langevin stochastic
equation that describes the evolution of the interface, and solved it
up to second order by numerical integration for finite system sizes,
recovering Eq.~(\ref{ws_ana}).

In this letter we mainly consider SF networks with $\lambda <3$
because they are representative of abundant systems in nature. We find
that, for the SRM model, $W_s$ has a crossover from a logarithmic
regime to constant regime, at a characteristic value of $N$ that
depends on the topology of the network. Also, we find that, for system
sizes above this characteristic size, $W_s$ increases as $\lambda$
decreases.

\vspace{1cm}

By stochastic numerical simulations, we study the SRM process on SF
networks. To generate the network, we use the Molloy Reed algorithm
(MR) or configurational model \cite{Molloy} and we use a minimum
degree $m=2$ because for this value we have a high probability of
obtaining only one component and thus a single interface
\cite{Cohen}. As for the maximum degree, the network has a natural
cut-off given by $k_{c} \sim N^{\frac{1}{\lambda -1}}$ \cite{Molloy}
and no structural cut-off is imposed, although this alternative will
be discussed later. We define a scalar field $h$ on the network, which
represents the system's feature we want to study, so that each node is
assigned a height $h_i$, with $i=1,...,N$. At $t=0$, we allocate the
nodes with a random height between $0$ and $1$. This initial condition
does not affect the scaling behavior of the roughness in its steady
state. At each time step, we deposit a particle on a node randomly
selected with a probability $1/N$. Denoting the selected node by $i$
and the set of its $k_i$ neighbors by $v_i$, we simulate the SRM
process according to the following rules \\
\noindent 1) if $h_i < h_j \, \forall \, j \,\epsilon \, v_i \Rightarrow h_i=h_i+1$ \, ,\\    
2) else, $l \in v_i$ : $h_l< h_i$ and $h_l<h_j \, \forall \, j \neq
l$, \, $j \,\epsilon \, v_i \Rightarrow h_l=h_l+1 $\, ,\\
\noindent and compute the roughness at the saturation as a function of
the system size.  

\begin{figure}[h] 
\vspace{1cm}
\begin{center}
\includegraphics[width=0.4\textwidth]{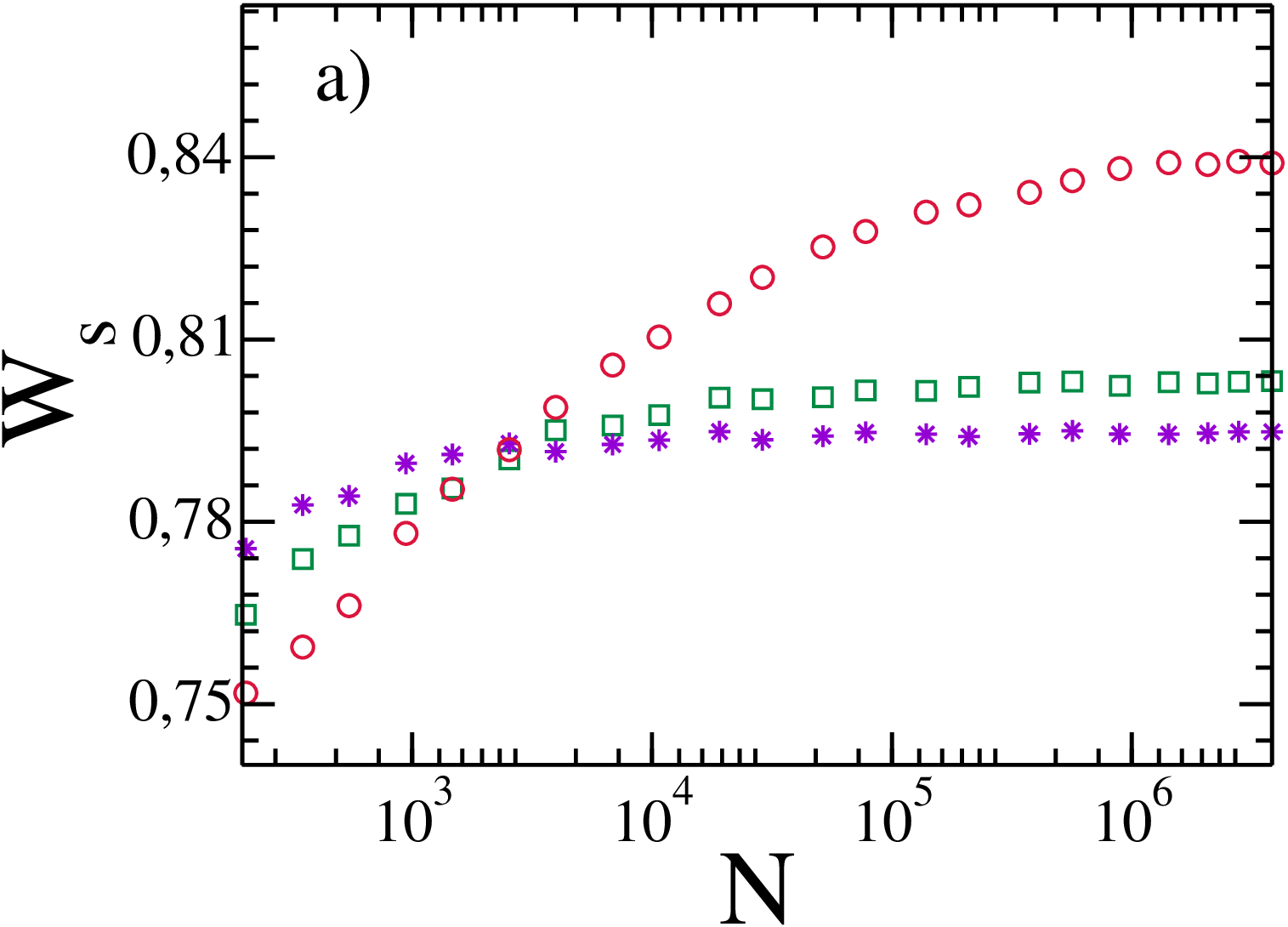}  
\includegraphics[width=0.4\textwidth]{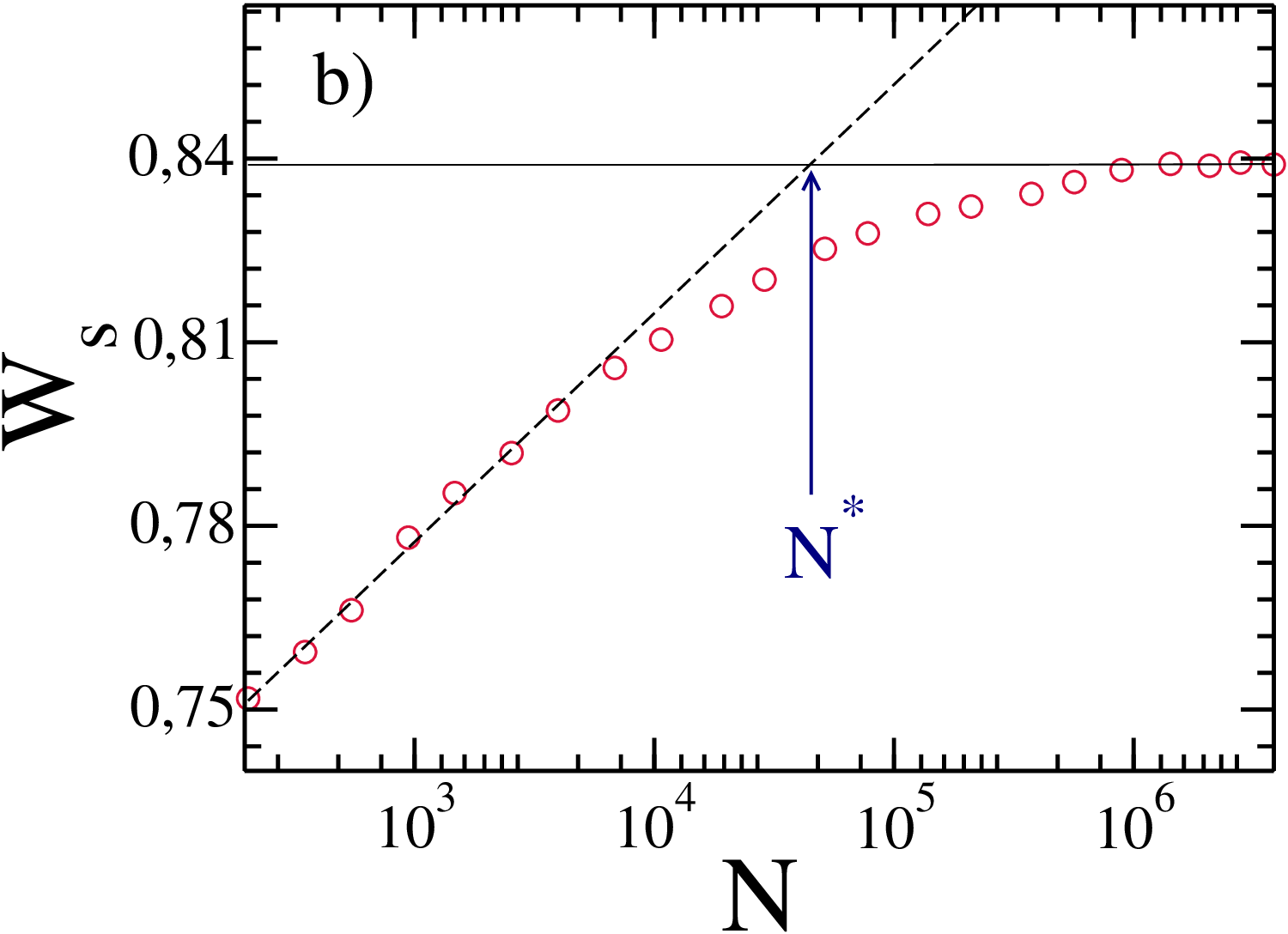}   
\end{center}
\caption{$a)$ $W_{s}$ as a function of $N$ in a linear-log scale, with
  $\lambda=2.6$ \,($\bigcirc$), $2.8$ \,($\Box$), $3.0$ \,($\ast$). We
  observe two regimes so that $W_s$ behaves as $\sim \, ln \, N$ for
  $N$ smaller than a certain characteristic size $N^{*}$, and it
  increases at a slower rate for $N>N^{*}$, finally becoming constant
  for a sufficiently large $N$. $b)$ $W_s$ as a function of $N$ in a
  linear-log scale for $\lambda=2.6$. It can be seen how we obtain the
  characteristic system size $N^{*}$ for the intersection of the two
  regimes.}\label{figura_1}
\end{figure}

In Fig. \ref{figura_1} (a) we plot the fluctuations
$W_s$ as a function of $N$ for different values of $\lambda$. We can
see that for $\lambda =3$ we obtain the behavior predicted by
Eq.~(\ref{ws_ana}), i.e, the fluctuations go rapidly to a constant when
the system size increases. For $\lambda <3$ and for a range of system
sizes ($N \lesssim 10^3$), the behavior of $W_s$ is logarithmic, which
agrees with Eq.~(\ref{ws_ana}), but when $N$ increases, the
fluctuations increase slower than a logarithmic, reaching a constant
that is independent of $N$ and only depends on $\lambda$. The scaling
behavior of $W_s$ suggests that above a certain system size $N=N^{*}$
(which depends only on $\lambda$), the fluctuations become independent
of $N$ and therefore the system has the same degree of
synchronization. It is worth noticing that the second regime was not
seen in \cite{Ana} since in their research they simulated systems
smaller than $N^{*}$. 
\begin{figure}[h]
\begin{center}
\vspace{1cm}
\includegraphics[width=0.4\textwidth]{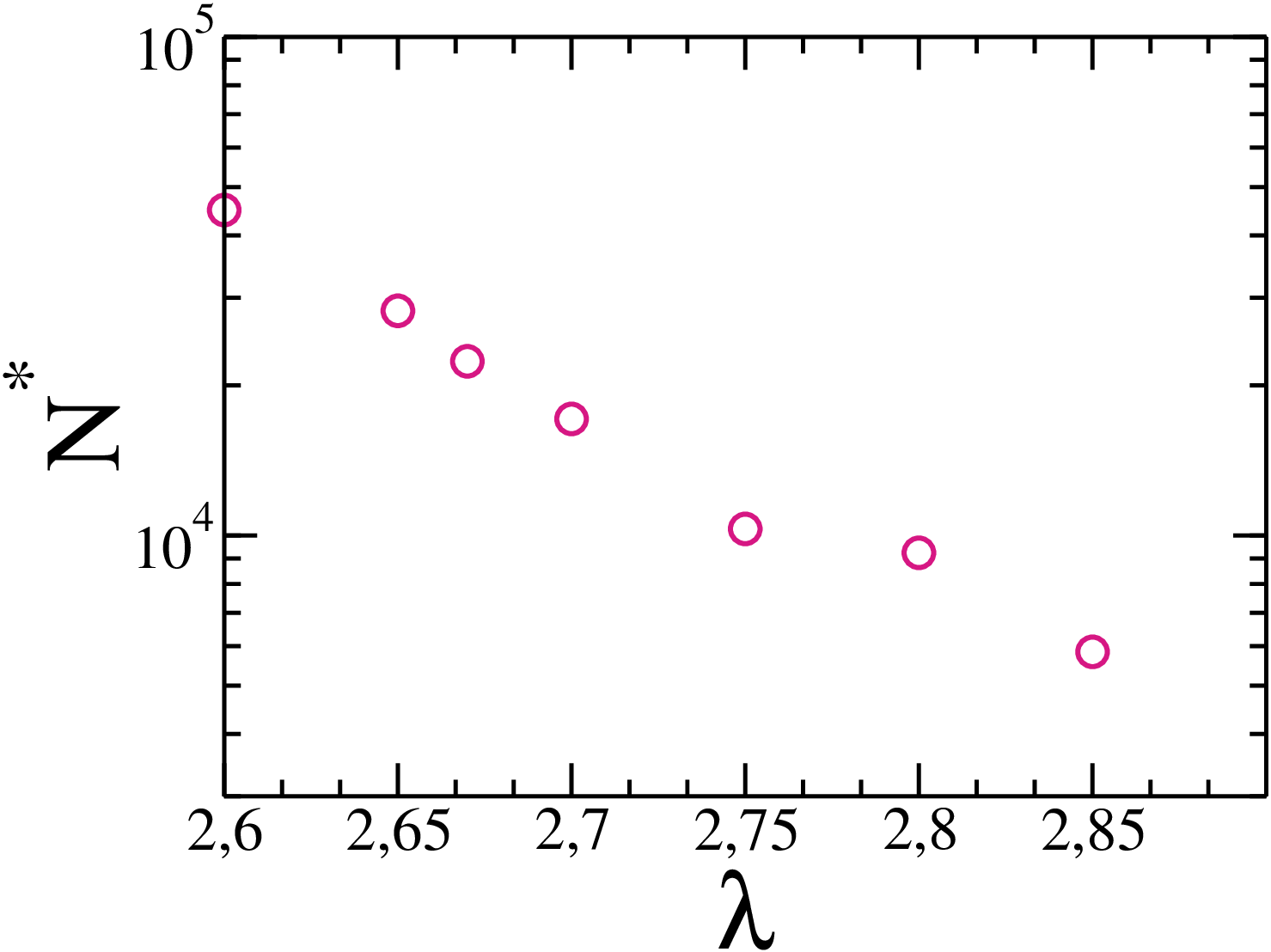}
\caption{$N^{*}$ as a function of $\lambda$, for SF networks with $2.6
  \le \lambda < 2.9$ with $m=2$.}\label{figura_2}
\end{center}
\end{figure}

We then estimate the system size for which the behavior of the
fluctuations changes from a logarithmic regime to a constant, i.e.,
the crossover between these two regimes. In Fig. \ref{figura_1} (b) we
plot $W_{s}$ as a function of $N$ only for $\lambda=2.6$ in order to
show how $N^{*}$ is determined. We compute $N^{*}$ for different
values of $\lambda$, and we see that $N^{*}$ decreases with $\lambda$
for $\lambda <3$, and $N^{*} \rightarrow 0$ for $\lambda \geq 3$, as
can be seen in Fig. \ref{figura_2}.

The behavior of $W_s$ with $N$ can be described as follows
\begin{eqnarray}
W_s \sim \left \{\begin{array}{ll}
  b\;ln(N) & \textrm{, for $ N<N^{*} $\,;}\nonumber\\
  W_{s}^{\infty} & \textrm{, for $N>N^{*}$\,,}\nonumber
\end{array}\right.
\end{eqnarray}
where $b \equiv b(\lambda)$ and $W_{s}^{\infty} \equiv
W_{s}^{\infty}(\lambda)$ is the roughness value in the thermodynamic
limit (above $N^{*}$). Thus we propose a scaling function $f(N/N^{*})$
where
\begin{equation}\label{f_escaleo}
f(N/N^{*})  \sim \left \{\begin{array}{ll}
  ln(N/N^{*}) & \textrm{, for $ N/N^{*}<1 $\,;}\\
  0 & \textrm{, for $N/N^{*}>1$\,.}
\end{array}\right.
\end{equation}  
Then, the behavior of $W_s$ for all the values of $N$ can be expressed as
\begin{eqnarray}
W_s=W_{s}^{\infty} + b \ f(N/N^{*}) \ .\nonumber
 \end{eqnarray}
To lose all dependency with $\lambda$, we work with the expression
$({W_s}^{\infty}-W_s)/b$ so that
\begin{eqnarray}
(W_{s}^{\infty} - W_{s})/b \sim -f(N/N^{*}) \sim \left  \{\begin{array}{ll}
  -ln(N/N^{*}) & \textrm{, for $ N/N^{*}<1 $ \, ;}\\
  0 & \textrm{, for $N/N^{*}>1$ \ .\nonumber}
\end{array}\right. 
\end{eqnarray}

\begin{figure}[h]
\begin{center}
\vspace{1cm}
\includegraphics[width=0.5\textwidth]{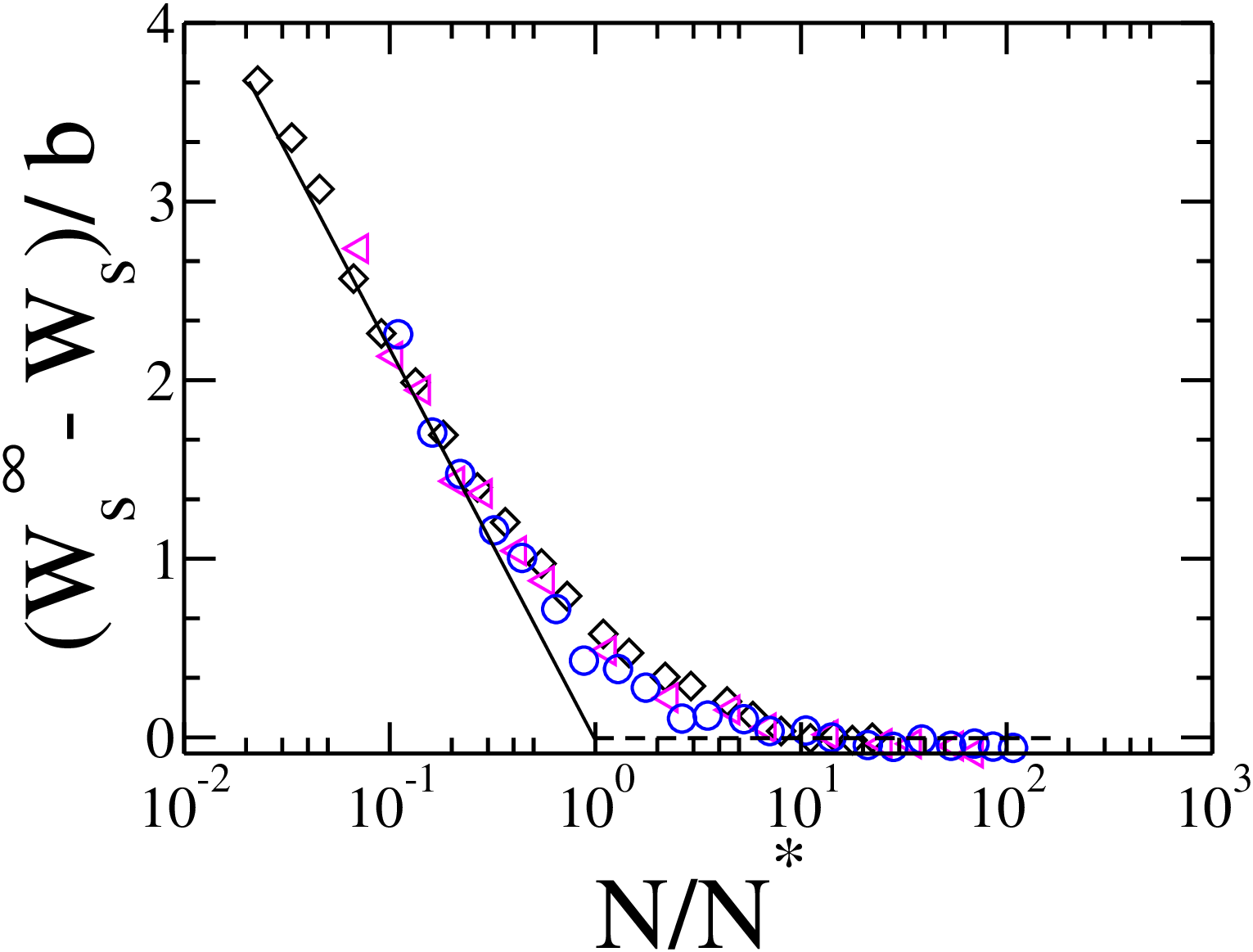}
\caption{$(W_{s}^{\infty} - W_{s})/b$ as a function of $N/N^{*}$ in a
  linear-log scale for $\lambda =2.6 (\diamond), 2.7 (\lhd)$ and $2.8
  (\bigcirc)$. The curves overlap, which proves the scaling hypothesis
  given by Eq.~(\ref{f_escaleo}) correct.}\label{figura_4}
\end{center}
\end{figure}

In Fig. \ref{figura_4} we plot $(W_s^{\infty}-W_s)/b$ as a function of
$N/N^{*}$. From the plot we can see that the curves indeed overlap,
which shows that our scaling hypothesis is correct.

\begin{figure}[h]
\begin{center}
\vspace{1.3cm}
\includegraphics[width=0.5\textwidth]{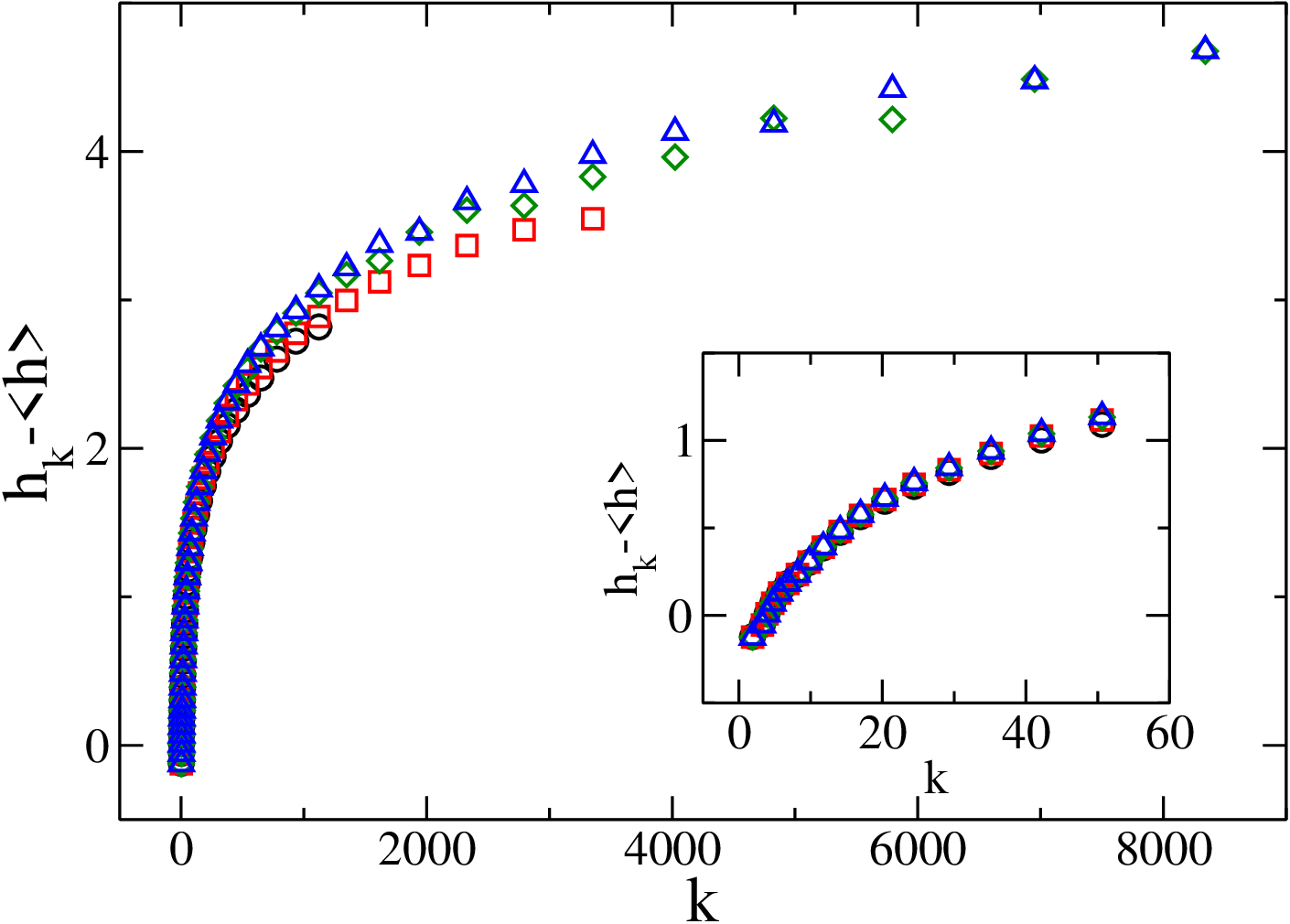}   
\caption{$h_{k}-\langle h \rangle$ as a function of $k$ for $N=6044 \,
  (\bigcirc), 32768 \, (\Box), 500000 \, (\diamond)$ and $1000000 \,
  (\triangle)$, and $\lambda=2.6$. The inset is an enlargement for
  small values of $k$; it can be seen that $h_{k}-\langle h \rangle$
  does not depend on $N$ in this region.}\label{figura_7}
\end{center}
\end{figure}

To understand the dynamics of the system as $N$ increases, we study
the behavior of the fluctuations relative to the topology of the
network, specifically the degree of the nodes. We compute the mean
height of the nodes with degree $k$, denoted by $h_{k}$. In
Fig. \ref{figura_7} we show $h_{k}-\langle h \rangle$ as a function of
$k$ for $\lambda =2.6$ and different values of $N$, below and above
$N^{*}$. We see that, in average, the difference between the height of
the nodes and the mean value of the entire network, increases with
$k$. Thus, for the SRM model, nodes with high degree worsen
synchronization while small degree nodes improve it. This is because
the nodes with high degree receive the excess of particles of their
low degree neighbors, which are the majority in SF networks. In order
to understand the effect of the high degree nodes on the behavior of
$W_s$, we study the SRM process for a network with a structural
cut-off of $k_{s} \sim N^{1/2}$ \cite{Boguna,Catanzaro}, which is
smaller than $k_{c}$ for $\lambda<3$ (not shown here). For this case,
we also find a crossover between two regimes at a characteristic size,
although it is larger than the one found with no structural
cut-off. This means that the hubs contribute to the finite size
effects for $N<<N^{*}$ with $\lambda < 3$.

We also notice that the behavior of $h_{k}-\langle h \rangle$ does not
depend on $N$ for small $k$, which means that low degree nodes do not
contribute to finite size effects on $W_{s}$. However, for larger $k$,
$h_{k}-\langle h \rangle$ increases with $N$. As $k$ increases, the
rate of increase of $h_{k}-\langle h \rangle$ decreases, so that the
nodes have heights more similar to one another. This behavior combined
with the fact that the probability of high connectivities is very low
determines that nodes with high degree do not contribute to an
increase in the fluctuations for $N \gg N^{*}$. Due to this combined
effect $W_s$ approaches to a constant for $N>>N^{*}$. On the other
hand, as $\lambda$ increases, the ratio of small degree nodes to hubs
increases. This explains why, around the value of $N^{*}$, the
synchronization improves as $\lambda$ increases. This can be seen from
Fig. \ref{figura_1}, where $N^{*}$ decreases as $\lambda \rightarrow
3$ and, consequently, the constant value to which $W_s$ approaches
gets smaller, which means that the system synchronizes better for
increasing $\lambda$.

To summarize, we studied the behavior of the fluctuations $W_s$ with
the system size $N$ in the steady state for the SRM model in SF
networks with $2.5 < \lambda < 3$, aiming to understand the role of
finite size effects in the system's synchronization. We found a
crossover between two different regimes at a characteristic size
$N^{*}$ below which $W_s$ has a logarithmic dependence with $N$ and
above which $W_s$ is constant. We measured $N^{*}$ for different
values of $\lambda$ and found that $N^{*}$ decreases with
$\lambda$. We also found that the synchronization enhances as
$\lambda$ increases. The behavior of $W_s$ with $N$ and $\lambda$ is
determined by the performance of high degree nodes in the dynamics of
the system, which reach heights above the mean value and worsen the
synchronization. However, the rate of increase of the heights compared
with the average height decreases with $k$ and, given that the degree
distribution also decreases with $k$, we conclude that high degree
nodes do not contribute to an increase of $W_s$ for $N \gg N^{*}$. It
is important to mention that eventhough high degree nodes are the
responsible of the finite size effects observed for $N<N^*$, the
explanation of the logarithmic behavior of $W_s$ goes beyond the aim
of our actual research. However, this will be the scope of future
researches. As for the heterogeneity of the network, as $\lambda$
increases, the proportion of high degree nodes decreases and the
previous effect is noted for smaller system sizes. In the limit
$\lambda \rightarrow 3$, $N^{*}\rightarrow 0$ and $W_s$ is constant
for all $N$.

\acknowledgments

We acknowledge UNMdP, FONCyT (Pict 0429/2013) and CONICET (PIP
00443/2014) for financial support. CELR is a acknowledges CONICET for
financial support.

\end{document}